\documentclass{amsart}
\begin{document}
\title{Weakly nonlocal irreversible thermodynamics -
the Guyer-Krumhansl and the Cahn-Hilliard equations}
\address{
Budapest University of Technology and Economics\\
Department of Chemical Physics\\
1521 Budapest, Budafoki \'ut 8.} \email{vpet@phyndi.fke.bme.hu}



\begin{abstract}
Examples of irreversible thermodynamic theory of nonlocal
phenomena are given, based on generalized entropy current.
Thermodynamic currents and forces are identified to derive the
Guyer-Krumhansl and Cahn-Hilliard equations. In the latter case
Gurtin's rate dependent additional term is received through the
thermodynamic approach.
\end{abstract}

\maketitle

\section{Introduction}

In the last decades there is a continuous interest in generalized
classical continuum theories to include  nonlocal phenomena. The
best treated and most popular type of nonlocal theories includes
higher- and higher order space derivatives (gradients) of the
relevant variables into the governing equations. Those
investigations are frequently called as gradient or weakly
nonlocal continuum theories. However, one cannot include higher
order gradients in the governing equations at an arbitrary place,
all equations are to be compatible with basic physical principles,
first of all with the Second Law of thermodynamics. More properly,
we are to construct weakly nonlocal differential equations where
the constitutive properties are determined so that solutions of
the governing dynamic equations result in nonnegative entropy
production.

Weakly nonlocal theories, related thermodynamics, are either
supposing a gradient dependent entropy function
\cite{Fal92a,PenFif90a}, or are considering a generalized form of
the entropy current \cite{MulRug96b,LebAta97a,MarAug98a}. However,
none of them uses the traditional methods of non-equilibrium
thermodynamics, there are no thermodynamic forces and currents
identified and no conductivity relations defined. In this paper we
will find suitable thermodynamic forces and currents with the help
of the generalized entropy current first used by Ny\'\i{}ri
\cite{Nyi86a}. Moreover, we will show that in case of heat
conduction the classical Onsagerian method leads to the well-known
weakly nonlocal generalization of the heat conduction equation
called Guyer-Krumhansl equation. On the other hand, we can get an
equation of the Cahn-Hilliard-type as an evolution equation of an
extensive variable in the presence of an internal variable of
nonlocal type. Several previous ad-hoc assumptions are explained
by constitutive relations in the present treatment (e.g. the
balance form equation of the heat current).

\subsection{Weakly nonlocal extended heat conduction}

When treating heat conduction phenomena in solid material, the
starting point is the balance equation of internal energy
\begin{equation}
\partial_t{u} + \nabla\cdot {\bf j}_q = 0.
\label{BalIE}\end{equation}

Here $u$ is the internal energy density and ${\bf j}_q$ is the
heat current density and $\partial_t$ denotes the partial time
derivative. Extended thermodynamics considers the heat current
density as an independent variable, introducing it into the basic
state space and a non-equilibrium entropy is defined as a function
on this extended state space. With these assumptions one can
explain inertial, memory effects getting an appropriate
phenomenological model of the wave heat conduction phenomena. The
key to receive a well tractable model is to identify suitable
thermodynamic currents and forces for a reasonable constitutive
theory. This can be achieved by supposing that the non-equilibrium
entropy function depends on the currents in a special way, as it
was first suggested by Gyarmati \cite{Gya77a}
\begin{equation}
s(u,{\bf j}_q) = s_0(u) - \frac{1}{2} {\bf j}_q\cdot {\bf m}
\cdot{\bf j}_q. \label{WaveEntropy}\end{equation}

Here {\bf m} is the so called matrix of the {\em thermodynamic
inductivities}. Further on we will assume that {\bf m} is constant
(as usual). The reciprocal absolute temperature is introduced as
the derivative of the entropy function according to the internal
energy $\frac{\partial s}{\partial u} = \frac{{\rm d} s_0}{\rm d
u} = 1/T$.

A natural physical assumption is that the entropy, being connected
to internal energy, does not flow without the heat flow.
Therefore, we accept, that the entropy current as a function of
our given state variables $(u,{\bf j}_q)$ has the property ${\bf
j}_s(u,{\bf 0})={\bf 0}$. Using Lagrange mean value theorem and
considering this condition, one can get the next functional form
\begin{equation}
{\bf j}_s(u,{\bf j}_q) = {\bf B}(u,{\bf j}_q)\cdot {\bf j}_q,
\label{EntrCurr}\end{equation}

\noindent where the {\bf B} called {\em current intensity factor}
can be supposed to be continuous at a neighborhood of the local
equilibrium state ${\bf j}_q = {\bf 0}$. One can observe that {\bf
B} is one tensorial order higher than the corresponding current.
After a short calculation, a consequence of our assumptions
(\ref{BalIE}), (\ref{WaveEntropy}) and (\ref{EntrCurr}) is, that
the entropy production has the following form
\begin{equation}
\sigma_s = \partial_t{s} + \nabla\cdot {\bf j}_s = ({\bf B} - 1/T
{\bf I}):\nabla\circ {\bf j}_q + (\nabla\cdot {\bf B} - {\bf
m}\cdot
\partial_t{\bf j}_q )\cdot {\bf j}_q.
\label{EntrProd}\end{equation}

Here {\bf I} is the second order identity tensor, $:$ and $\circ$
are the notations of the double contraction and the tensorial
product, respectively. The entropy production is nonnegative and
equals zero only in thermodynamic equilibrium according to the
Second Law.

We are looking for constitutive relations for {\bf B} and
$\partial_t{\bf j}_q$ that automatically fulfill the inequality
above. The following force-current structure can be chosen

\vskip 3mm
\begin{center}\begin{tabular}{ccc}
 \hfill\vline & Local & Nonlocal \\
\hline \hfill\vline \\
Force  \hfill\vline & $\nabla\cdot {\bf B} - {\bf m} \cdot
\partial_t {\bf
    j}_q$ & $\nabla\circ {\bf j}_q$\\
Current \hfill\vline & ${\bf j}_q$ & ${\bf B} - \frac{1}{T}
\bf{I}$
\end{tabular}.\end{center}\vskip 3mm

The linear approximation of Onsager can be written as follows:
\begin{eqnarray}
{\bf j}_q &=&
    {\bf L}_{11} (\nabla\cdot {\bf B} -
        {\bf m}\cdot \partial_t{\bf j}_q)
    + {\bf L}_{12}\nabla\circ {\bf j}_q, \label{O-1}\nonumber\\
{\bf B} - \frac{1}{T} \bf{I} &=&
     {\bf L}_{21}(\nabla\cdot {\bf B} -
        {\bf m} \cdot\partial_t{\bf j}_q)
    + {\bf L}_{22} \nabla\circ {\bf j}_q.
\label{Ons2}
\end{eqnarray}

Of course we could give more general relations among the
thermodynamic currents and forces, but this linear approximation
seems to be quite general for our further purposes. From the
equations above {\bf B} can be easily eliminated and we can give a
general constitutive relation purely for ${\bf j}_q$
\begin{eqnarray}
{\bf j}_q &=& {\bf L}_{11} (\nabla\frac{1}{T(u)}
    - {\bf m} \cdot \partial_t{\bf j}_q)
+ {\bf L}_{11} \nabla\cdot\left( ({\bf L}_{22}
    - {\bf L}_{21} {\bf L}_{11}^{-1}
    {\bf L}_{12})\nabla\circ {\bf j}_q \right.\nonumber\\
    &+& \left. {\bf L}_{21} {\bf L}_{11}^{-1}  {\bf j}_q \right)
+ {\bf L}_{12} \nabla\circ {\bf j}_q.
\label{genco4nst}\end{eqnarray}

This general relation gives the first approximation considering
the entropy triggering effects of nonlocalities. The relations
(\ref{genco4nst}) and (\ref{BalIE}) together result in the system
of equations to be solved for $u$ and ${\bf j}_q$.

Let us now restrict ourselves to the isotropic case. According to
the Curie principle (representation theorems of isotropic
tensors), there is no cross coupling in the conductivity
equations. In a strictly linear approximation (when ${\bf L}$ is
constant) we get ${\bf m}= m {\bf I}$, ${\bf L}_{11}= l {\bf I}$,
where $m,l$ are a constant scalars and ${\bf L}_{22}$ can be
written with indices as:
$$
({L}_{22})_{ijkl} = l_1\delta_{ik}\delta_{jl} +
l_2\delta_{il}\delta_{jk} + l_3 \delta_{ij}\delta_{kl},
$$

\noindent Here $l_1$, $l_2$, $l_3$ and $l$ are positive constant
scalar material parameters to ensure nonnegative entropy
production. $m$ is also positive supposing concave entropy in the
extended state space. Therefore the linear approximation of
Onsager (\ref{Ons2}) reduces to
\begin{eqnarray}
{\bf j}_q &=& l (\nabla\cdot {\bf B} -  m \partial_t{\bf j}_q),\nonumber\\
{\bf B} - \frac{1}{T} \bf{I} &=& l_1 (\nabla \circ {\bf j}_q) +
l_2(\nabla \circ {\bf j}_q)^* + l_3 \nabla \cdot{\bf j}_q\bf{I},
\label{HeatO-2}\end{eqnarray}

\noindent where $*$ denotes the transpose. Using this expression
we can get
\begin{equation}
l m \partial_t{\bf j}_q + {\bf j}_q =
    l \nabla \frac{1}{T} + l (l_1 \Delta {\bf j}_q +
    (l_2+l_3)\nabla\circ\nabla\cdot{\bf j}_q).
\label{IsoO-1}\end{equation}

This is exactly the Guyer-Krumhansl equation for the heat current,
introduced to describe the thermal properties of some crystals at
low temperatures \cite{GuyKru66a1}. Originally it was derived with
the use of kinetic physics but other derivations are based on
two-fluid hydrodynamics \cite{Enz74a}. In the last decades there
were several attempts to get these equations from pure
non-equilibrium thermodynamics
\cite{LebDau90a,Net93a,LebGre96a,LebAta97a, LebAta98a}. However,
all of these derivation contain several ad-hoc assumptions as
supposing a balance-like dynamics for the heat current ${\bf j}_q$
and specially simplified constitutive equations. Moreover, the
derivations mentioned above does not have the heuristic power of
irreversible thermodynamics, therefore the generalization of their
assumptions in case of more difficult situations can be very
ponderous if not impossible.


In the special case of $l_1=l_2=l_3=0$ we can get the
Cattaneo-Vernote equation for the heat current. In this case the
entropy current has its traditional form ${\bf j}_s = {\bf
j}_q/T$. Further, we get the Fourier heat conduction equation when
${\bf m}={\bf 0}$, in the case of local equilibrium. Assuming the
caloric state function in the form $u = c T$, the weakly nonlocal
extended heat conduction equation can be given directly for the
temperature as
\begin{equation}
l m c\partial_{tt}T + c\partial_t{T} + l \Delta \frac{1}{T} + l
c(l_1+l_2+l_3) \Delta\partial_t{T} =0. \label{GuKru}\end{equation}

Let us remark here that several generalizations are known for the
classical entropy current expression from the investigations in
extended thermodynamics \cite{MulRug96b} to the multifield
theories that include finite length localization instabilities in
damage mechanics models \cite{MarAug98a}. The treated Ny\'\i{}\-ri
form of the entropy current makes possible to exploit the Second
Law in an Onsagerian spirit.

Another important remark is, that the heat current ${\bf j}_q$ and
the intensity factor ${\bf B}$ are internal variables in
thermodynamic sense. If we do not eliminate ${\bf B}$ from the
material equations, then it can be regarded as a current density
of the heat current ${\bf j}_q$ \cite{Ver83a},  because the
dynamic equation for the heat current derived above turns out to
have a special balance form (more properly it will be an equation
of Ginzburg-Landau type). Supposing that both the entropy and the
entropy current depend on {\bf B}, we can continue to introduce
internal variables of nonlocal type in the following way: we
construct a new current with the help of the given current
intensity factor and introduce a new current intensity factor of
{\bf B} in the entropy current. In this way we can get a
hierarchical approximation scheme of nonlocal phenomena based on
thermodynamic considerations.

\section{Weakly nonlocal extensive variable - the Cahn-Hilliard
equation}

Now, another example will be shown that similar structural
assumptions of the entropy current can help to build up a
current-force system in more general cases, too. Let us consider a
continuum with a single scalar extensive variable $a$ (e.g. mass
density). Therefore a balance equation is given for $a$
\begin{equation}
\partial_t{a} + \nabla\cdot {\bf j}_a = 0.
\label{BalEV}\end{equation}

However, as we do not consider a simple local equilibrium
approximation, we further extend the state space of extended
thermodynamics and introduce an internal variable
$\boldsymbol{\xi}$ with the tensorial order of the current
density. In this case our basic state space is spanned by the
variables $(a, {\bf j}_a, \boldsymbol{\xi})$. We are to
investigate the additional nonlocal terms in the governing
equations, therefore, contrary to the previous example, we will
neglect the memory effects, so the entropy function will depend
only on the extensive quantities
\begin{equation}
s(a,{\bf j}_a,\boldsymbol{\xi}) = s_0(a).
\label{CHEntropy}\end{equation}

The corresponding intensive quantity is denoted by $\Gamma(a) =
\frac{{\rm d} s_0}{\rm d a}$. Let us investigate now the
functional form of the entropy current ${\bf j}_s(a, {\bf j}_a,
\boldsymbol{\xi})$. There are some natural physical assumptions
that we apply:
\begin{itemize}
\item There is no entropy flow in the absence of the flow of $a$
and with zero internal variable:
$$
{\bf j}_s(a, {\bf 0},  {\bf 0}) = {\bf 0}.
$$

\item In case of zero internal variable the entropy flow reduces
to the classical form:
$$
{\bf j}_s(a, {\bf j}_a, {\bf 0}) =
    \Gamma {\bf j}_a.
$$
\end{itemize}

Therefore, according to Lagrange's mean value theorem and the
mentioned natural physical assumptions we may write that
\begin{equation*}
{\bf j}_s(a,{\bf j}_a, \boldsymbol{\xi}) =  \Gamma {\bf j}_a +
    {\bf A}\cdot \boldsymbol{\xi}.
\end{equation*}

\noindent where {\bf A} is a continuous constitutive function on
the state space and its tensorial order is one order higher than
the order of $\boldsymbol{\xi}$. We may observe that in this case
the current intensity factor {\bf B} connected to the heat
current is replaced by its local equilibrium value $\Gamma$
according to our assumption with neglecting the memory effects
(expressed by the condition (\ref{CHEntropy})). That simplifying
assumptions on the entropy and the entropy flow yields the
entropy production as
$$
\sigma_s = \Gamma\partial_t{a} + \nabla\cdot(\Gamma {\bf j}_a +
{\bf A}\cdot\boldsymbol{\xi})=
    {\bf j}_a \cdot \nabla\Gamma +
    {\bf A}:\nabla\circ\boldsymbol{\xi} +
    (\nabla\cdot{\bf A}) \cdot \boldsymbol{\xi}.
$$

Let us consider now the strictly linear approximation with
constant coefficients in case of isotropic materials only. Now the
choice of the currents and forces in straightforward. There are
two coupled terms in the linear laws
\begin{eqnarray}
{\bf j}_a &=& l_{11} \nabla \Gamma + l_{12} \nabla \cdot{\bf A},
\label{CHO-1}\\
\boldsymbol{\xi} &=& l_{21} \nabla\Gamma + l_{22} \nabla \cdot
{\bf A},
\label{CHO-2}\\
{\bf A} &=& l_1 (\nabla \circ \boldsymbol{\xi}) + l_2(\nabla \circ
\boldsymbol{\xi})^* + l_3 \nabla \cdot\boldsymbol{\xi} \bf{I}.
\label{CHO-3}
\end{eqnarray}

Here the matrix of the conductivity coefficients is positive
definite to ensure nonnegative entropy production. We can
eliminate $\boldsymbol{\xi}$ from the first two equations:
$$
\boldsymbol{\xi} = l_{22} l_{12}^{-1} {\bf j}_a + (l_{21} - l_{22}
l_{12}^{-1} l_{11})\nabla \Gamma.
$$

Eliminating {\bf A} and $\boldsymbol{\xi}$ from the first
equation, we get the following equality
\begin{equation}
{\bf j}_a = l_{11}\nabla(\Gamma - \alpha \Delta\Gamma) +
    l_1 l_{22} \Delta {\bf j}_a +
    l_{22}(l_2+l_3) \nabla\circ\nabla\cdot {\bf j}_a,
\label{jeq}\end{equation}

\noindent where $\alpha = (l_1+l_2+l_3)(l_{11}l_{22} -
l_{12}l_{21})$ is positive.

So we eliminated the internal variable and received (\ref{BalEV})
and (\ref{jeq}) as dynamic equations to be solved.

Putting (\ref{jeq}) into the balance (\ref{BalEV}) we finally
obtain
\begin{equation}
\partial_t{a}+ l_{11}\Delta(\Gamma - \alpha \Delta \Gamma) -
l_{22}(l_1+l_2+l_3) \Delta \partial_t{a}= 0.
\label{Cahn-Hilliard}\end{equation}

This equation is astonishingly similar to the Cahn-Hilliard
equation, but there are some essential differences.
\begin{itemize}
\item The last extra term does not appear in the traditional equation.
Gurtin received a similar additional term with the help of
microforce balance (a strong additional assumption) in the special
case when the variable $a$ is the mass density \cite{Gur96a}.
Supposing that $l_{22}=0$, this additional term vanishes. Let us
remark that this rate dependent term appears in the linear
Onsagerian conductivity equations at the same thermodynamic
approximation level as all the others.
\item Under the second Laplace operator there is the intensive
variable $\Gamma$, instead of the extensive $a$, contrary to the
original Cahn-Hilliard equation.
\item All of the material parameters in (\ref{Cahn-Hilliard}) are
positive, as a consequence of the Second Law, a nonnegative
entropy production. In the original equation the signs of the
coefficients are fixed according to stability considerations that
are not connected to the entropy balance.
\item The nonlinearities and anisotropies
show a different structure than in the original equation.
\item In most of the previous derivations of the Cahn-Hilliard equations,
a "generalized chemical potential" appeared by analogy, and was
inserted without any further ado into the mass balance. Here in
our derivation we received a real generalization of the diffusion
current in the same sense and with the same method as in the
original irreversible thermodynamic approach.
\end{itemize}

\section{Discussion}

In this letter only some basic examples are treated. A more
general theory of weakly nonlocal irreversible thermodynamics can
be created according to the methods discussed above. It makes
possible to incorporate and investigate several other classical
nonlocal equations (phase field, Kardar-Parisi-Zhang, etc..) into
the well established structure of non-equilibrium thermodynamics
and investigate them form the point of view of structural
compatibility. The present treatment has several serious
advantages if compared to other approaches. First of all the
compatibility of those equations with the Second Law of
thermodynamics can be investigated. These researches can deepen
our understanding of the Second Law that seems to emerge as a
general theory of material stability in a static and also in a
dynamic sense \cite{Mat00a}.

As we have already mentioned, the main advantage of the above
scheme is the use of the force-current relations of Onsager, which
offers a consequent approximation method. It is straightforward
to go beyond the linear approximation and consider higher order
terms in the series. Furthermore, with this scheme we can be sure
that the amendment of the Second Law suggested by Muschik and
Ehrentraut \cite{MusEhr96a} is fulfilled, that is the positivity
of the entropy production is ensured by purely constitutive
assumptions.

One can say that the heuristic power of the Onsagerian approach
can be a disadvantage if contrasted with the more general
exploitations of the entropy inequality. However, an application
of the Liu procedure \cite{Liu72a,MusAta00a2} shows that our
assumptions on the entropy current are not only correct, but
essentially one cannot construct more general constitutive
relations that would satisfy the amendment of the Second Law.
These problems and some other example applications are to be
treated in a forthcoming paper \cite{Van01m2}.

Finally we remark, that there are not well established methods to
derive weakly nonlocal equations in kinetic physics (see e.g.
\cite{Lib90b,Net93a}). According to the present investigations,
the role of the entropy current can be essential to develop a
powerful approximation scheme, like the moment series expansion in
case of memory effects \cite{MulRug96b}.

\section{Acknowledgements}

This research was supported by OTKA T034715 and T034603.

\end{document}